\documentclass[prl,twocolumn, superscriptaddress]{revtex4}
\usepackage{graphicx,amsmath,amssymb,amsxtra}
\renewcommand{\narrowtext}{\begin{multicols}{2} \global\columnwidth20.5pc}

\def\be{\begin{eqnarray}}
\def\ee{\end{eqnarray}}

\newcommand{\Eq}[1]{Eq.~(\ref{#1})}
\newcommand{\Fig}[1]{Fig.~(\ref{#1})}

\newcommand{\cL}{{\cal L}}

\newcommand{\sgn}{\text{sgn}}

\begin{document}

\title{Momentum resolved tunneling into the Pfaffian and anti-Pfaffian edges}

\author{Alexander Seidel}
\affiliation{Department of Physics and Center for Materials Innovation, Washington University, St. Louis, MO 63136, USA}
\author{Kun Yang}
\affiliation{NHMFL and Department of Physics, Florida State
University, Tallahassee, FL 32306, USA}

\date{\today}

\begin{abstract}

We calculate the electron spectral functions at the edges of the Moore-Read Pfaffian and anti-Pfaffian fractional quantum Hall states, in the clean limit. We show that their qualitative differences can be probed using momentum resolved tunneling, thus providing a method to unambiguously distinguish which one is realized in the fractional quantum Hall state observed at filling factor $\nu=5/2$.
We further argue that edge reconstruction, which may be less important in the first excited Landau
level (LL) than in the lowest LL, can also be detected this way if present.

\end{abstract}

\maketitle

{\em Introduction.}
Fractional quantum Hall (FQH) systems represent one of the
richest and most fascinating classes of interacting electron
systems known to-date. Possible realizations may include
states supporting non-abelian statistics, which have been
proposed to allow fault-tolerant ``topological'' quantum computing\cite{kitaev, DFN}.
However, in general the striking transport properties that gave the FQH effect
its name are not sufficient to discriminate between various
classes of different states that may occur at a given Landau level (LL)
filling factor $\nu$.
The most hopeful experimental candidate system
for a non-abelian state is the FQH state
at $\nu=5/2$ \cite{willett1}. Possible non-abelian states
explaining the $\nu=5/2$ plateau include the Moore-Read ``Pfaffian'' (Pf)\cite{MR}
and its particle-hole conjugate counterpart, the ``anti-Pfaffian'' (AP) \cite{lee,levin}.
These two states have very closely related bulk properties
and most fundamentally differ through the physics of their edge states.
Recent experiments involving quasi-particle tunneling
between opposite edges across constrictions (or point contacts) have probed quasi-particle charge\cite{doblev, radu},
and may have revealed signatures of non-abelian statistics\cite{willett2}. They do not, however, allow for a clear distinction between the Pf and AP states; in fact only the experiment in Ref. \onlinecite{radu} is sensitive to the difference between these two states, which shows up as a {\em quantitative} difference in certain power-law exponents. In this paper we show that
momentum resolved electron tunneling (MRT) through a clean and extended junction \cite{kang, yang_i,huber1, huber2} into the edge of the
$\nu=5/2$ state gives rise to {\em qualitative} differences in the signals, and may thus be the most promising diagnostic tool
to distinguish these two states from one another, as well as from
other possible states.

\indent{\em Experimental setup and physical assumptions.}
\begin{figure}[t]
\begin{center}
\includegraphics[width=3.25in]{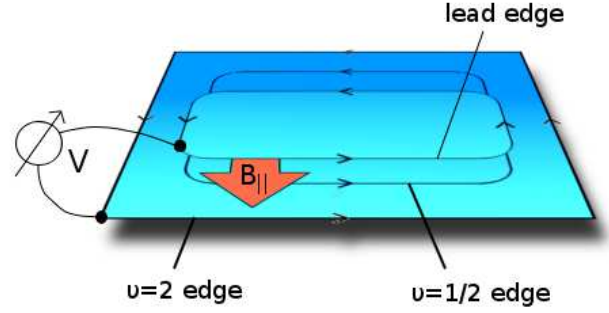}

  \caption{\label{setup}
(Color online). Schematic setup for momentum resolved tunneling.
A $\nu=1/2$ edge in the second Landau
level is contained within the outer $\nu=2$
edge of the filled lowest Landau level.
Tunneling takes place between the $\nu=1/2$
edge and a lead, consisting of the edge
of a narrow $\nu=1$ strip. By adjusting the
position of the strip and/or the in-plane
magnetic field $B_{||}$, tunneling into
a co-propagating (front) or a counter-propagating (back)
$\nu=1$ edge may be realized.
}
\end{center}
\end{figure}
A possible experimental setup
is depicted in \Fig{setup}, which is currently being pursued experimentally\cite{yacoby}.
The tunneling is between the $\nu=1/2$ edge in the
second LL and the edge of a Hall droplet
in a vertically separated layer,
which we assume to be in a $\nu=1$ state.
We will consider both co- and counter-propagation lead
geometries, i.e. the $\nu=1$ edge state propagates along the same or the opposite direction compared to that of the $\nu=5/2$ edge state.
Note that
the $\nu=1/2$ edge of the second LL will be contained well inside the
edge of the filled lowest LL of the $\nu=5/2$ droplet (see \Fig{setup}).
While this may complicate tunneling into this edge with other settings,
in that of \Fig{setup} this problem is circumvented
by positioning a narrow $\nu=1$ strip on top of
the $\nu=1/2$ edge.
This allows tunneling
into both a co-propagating as well as a counter-propagating
$\nu=1$ edge (see caption).


The Pfaffian edge theory consists of the sum of a massless chiral
fermion and massless chiral boson Lagrangian density with co-propagating
velocities, $\cL_{\text{Pf}}(\psi,\phi)=\cL_\psi+\cL_\phi$, where
\begin{subequations}
\begin{align}
 & \cL_\psi= i\psi\,(\partial_t+v_n\partial_x)\psi\\
 & \cL_\phi=\frac{1}{2\pi}\,\partial_x\phi\,(\partial_t+v_c\partial_x )\phi
\end{align}
\end{subequations}
and $v_n$ and $v_c$ are the neutral and the charged mode velocities, respectively.
Here, $v_n\ll v_c$ is expected due to the fact that $v_c$ is associated
with the larger Coulomb energy scale, in agreement with the numerics of Refs. \onlinecite{wan1,wan2}. In presenting the theory of the AP edge, we will follow Ref. \onlinecite{lee}, with the essential difference that
we assume disorder to be so weak that momentum remains a good quantum number at the length and energy scales relevant to the experiment, and do not include it.
On the other hand, disorder has been a key ingredient leading
to the conclusion of universal scaling dimensions in Ref. \onlinecite{lee}. Here we will argue that the same universal exponents are also obtained, to very good approximation, based on the separation of energy scales between charged and neutral modes.
In the spirit of Refs. \onlinecite{lee,levin}, we thus write the theory of the
AP edge as the sum of the Pfaffian edge Lagrangian with all mode velocities reversed and that of a $\nu=1$ edge, together with a density-density interaction
between the two charge modes:
\begin{equation}\label{LAP}
  \cL_{AP}=\frac{1}{4\pi}\partial_x\phi_1\,(\partial_t+v_1\partial_x )\phi_1
+\bar\cL_{\text{Pf}}(\psi,\phi_2)+\frac{v_{12}}{2\pi}\partial_x\phi_1\partial_x\phi_2\,.
\end{equation}
Here,
the field $\phi_1$ describes the $\nu=1$ edge, and
$\bar\cL_{\text{Pf}}$ denotes the Pfaffian Lagrangian discussed above
with the formal substitution $\partial_x\rightarrow -\partial_x$.
In \Eq{LAP}, the velocity parameters and the interaction $v_{12}$ are
independent, but their relative orders of magnitude are set by the dominance of
the Coulomb energy scale, as will become apparent shortly below.
To see this,
we carry out the charge/neutral decomposition of Ref.\cite{lee}
via $\phi_\rho=\phi_1-\phi_2$, $\phi_\sigma=\phi_1-2\phi_2$.
The physical significance of $\phi_\rho$ is that
$\rho_{\text{tot}}=-\partial_x\phi_\rho/2\pi$
is the total charge density at the edge, while $\phi_\sigma$ is the
linear combination of $\phi_1$ and $\phi_2$ that
commutes with $\rho_{\text{tot}}$.
In terms of the new fields,
\begin{equation}
\begin{split}\label{LAP2}
\cL_{AP}=&\frac{1}{2\pi}\partial_x\phi_\rho\,(\partial_t+v_\rho\partial_x )\phi_\rho
+
\frac{1}{4\pi}\partial_x\phi_\sigma\,(-\partial_t+v_\sigma\partial_x )\phi_\sigma
\\
&+\frac{v_{\rho\sigma}}{2\pi}\partial_x\phi_\rho\partial_x\phi_\sigma
+i\psi(\partial_t-v_n\partial_x)\psi
\,.
\end{split}
\end{equation}
where $v_\rho$, $v_\sigma$ and $v_{\rho\sigma}$ are
simple linear combinations of $v_1$, $v_c$ and $v_{12}$.
In \Eq{LAP2}, however, the large Coulomb energy scale should
enter only the coupling of the total charge density
with itself, i.e. $v_\rho$. All other coupling constants
are independent of this energy scale, and are expected to be
much smaller, i.e. $v_\sigma\sim v_{\rho\sigma}\ll v_{\rho}$.
Under these circumstances, the inter-mode coupling constant
$v_{\rho\sigma}$ has a very small effect of order $v_{\rho\sigma}/v_\rho$
on the scaling dimensions of operators.
To a good approximation, we may thus set $v_{\rho\sigma}\approx 0$,
which allows us to read the scaling dimensions of various
operators directly off of \Eq{LAP2}. Here we are only interested
in the most relevant operators that have the quantum numbers
of the electron operator. These operators and their scaling
dimension are then identical to those identified in
Refs.\onlinecite{lee, levin}. We emphasize, however,
that the
argument given here relies on the dominance of Coulomb interactions
only and does not invoke disorder, which played
a central role in Ref. \onlinecite{lee}.
As a result, the edge theory \Eq{LAP2} retains two distinct
counter-propagating neutral mode
velocities, $v_n$ and $v_\sigma$.\\
\indent{\em Electron operators and spectral functions.}
An electron operator of minimal scaling dimension $3/2$
is given by $\psi_{\text{el},1}(x) =\psi(x)\exp(-2i\phi_\rho(x))$,
for both the AP and Pf edge theory (we identify $\phi\equiv\phi_\rho$
in the latter).
In the Pf case, this is the unique leading electron operator,
whereas there are two more such operators of equal scaling dimension
in the AP case. These may be taken to be
$\psi_{\text{el},2,3}(x) =\exp(\pm i\phi_\sigma(x))\exp(-2i\phi_\rho(x))$.
The leading term in the electron operator at the AP edge is thus
a superposition of the operators $\psi_{\text{el},j}$, $j=1,2,3$. However,
all cross-correlations between different $\psi_{\text{el},j}$ vanish,
and the electron Green's function is of the form
$G(t,x)\simeq -i \sum_j a_j\langle \psi^\dagger _{\text{el},j}(t,x)\psi_{\text{el},j}(0,0)\rangle$. We will discuss the contributions to the electron spectral function of these correlators separately. Their real space structure is given by
\begin{equation}\label{corr}
\begin{split}
\langle \psi^\dagger _{\text{el},j}(t,x)\psi_{\text{el},j}(0,0)\rangle\propto&\\
\frac{i\, \sgn(u_n)}{x-u_nt+i0^+\sgn(u_nt)}&\,\frac{-1}{(x-u_ct+i0^+\sgn(t))^2}\,.
\end{split}
\end{equation}
In the above, $u_c$ equals $v_c$ in the Pf case and $v_\rho$ in the AP
case, whereas $u_n$ equals $v_n$ in the Pf case and $-v_n$ in the AP
case for $j=1$, and $-v_\sigma$ for $j=2,3$.
From \Eq{corr} one can obtain the Fourier transform
$G(\omega,q)$ of the electron Green's function, and
the spectral function $A(\omega,q)=-(1/\pi)\text{Im}G(\omega,q)\sgn(\omega)$.
More directly, $A(\omega,q)$ can be obtained from the convolution
method detailed in Ref. \onlinecite{MY1}.
For each of the leading contributions shown in \Eq{corr},
the result $A_j(\omega,q)$ is given by
\begin{equation}
A_j(\omega,q)\propto\Theta\left[u_n(\omega-qu_n)(qu_c-\omega)\right]\,\frac{|\omega-qu_n|}{(u_c-u_n)^2}
\end{equation}
with $\Theta$ the Heaviside step function.
\begin{figure}[t]
\begin{center}
\includegraphics[width=2.28in]{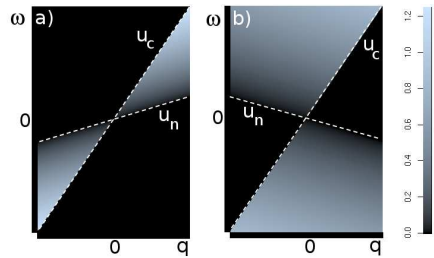}
  \caption{\label{spectral}(Color online). The electron spectral function
in the low-energy long-wavelength limit
for the Pfaffian (a) and anti-Pfaffian (b) edge.
$q$ is measured relative to the Fermi wavevector at the edge
for a specific electron operator.
Dashed lines indicate $\omega=u_cq$ and $\omega=u_nq$.
In b), only the contribution to the spectral function
due to one of three leading electron operators
at the anti-Pfaffian edge is shown (see text).
}
\end{center}
\end{figure}
\begin{figure}[!]
\begin{center}
\includegraphics[width=2.75in]{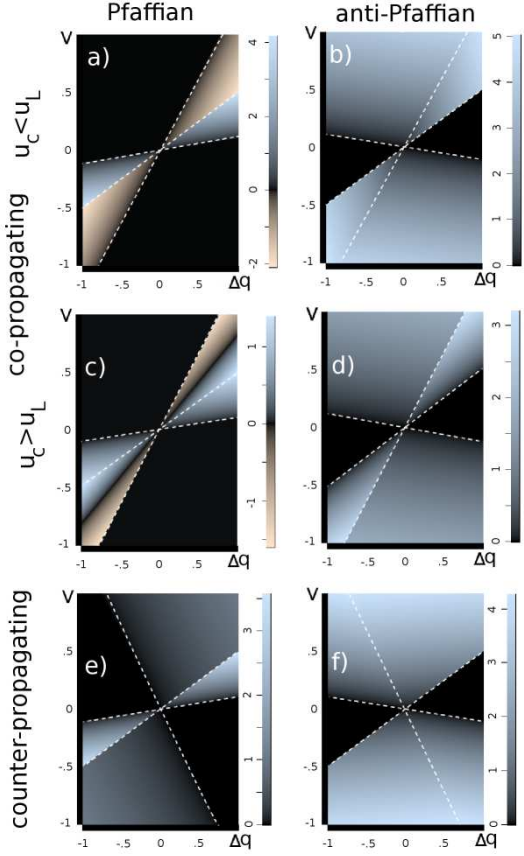}
  \caption{\label{result}(Color online).
$dI/dV$ as a function of applied voltage $V$ and wavenumber change
$\Delta q$ (all units arbitrary). 
$\Delta q$ is related to the in-plane magnetic field
$B_{||}$ via $\Delta q=ed(B_{||}-B_j)/\hbar c$ (see text).
The first column shows results for the Pfaffian case ($u_n>0$), the second column applies to the anti-Pfaffian case ($u_n<0$). The last row assumes tunneling
into a counter-propagating $\nu=1$ lead edge ($u_L<0$), the first two
rows assume a co-propagating lead edge ($u_L>0$),
with $u_L$ greater than (less than) $u_c$ in the first (second) row.
Dashed lines correspond to $eV=u_nq$, $eV=u_cq$, and $eV=u_Lq$, respectively,
 and mark the
boundaries of different regions across which $dI/dV$ and/or its derivatives
have discontinuities.
For clarity, we have chosen $|u_n|=0.1$ always, and $u_c=0.5$, $|u_L|=1.3$
for the first and last row, whereas $u_c=1.3$ and $u_L=0.5$
in the second row. The signs of $u_n$ and $u_L$ are varied as
appropriate to each case.
Distinctive features discriminating between the Pfaffian
and anti-Pfaffian cases are clearly visible.
In addition, the $dI/dV$ plots shown here for the anti-Pfaffian edge
take into account only one of three leading electron operators
for simplicity. In the full $dI/dV$ signal, each of these operators
makes a contribution of the kind shown above, but possibly with different
horizontal offsets, and with only two of three  neutral mode velocities
identical (see text).
}
\end{center}
\end{figure}
The results are plotted in \Fig{spectral} for both the Pf ($u_n>0$)
and the AP case ($u_n<0$). The
presence/lack of a counter-propagating mode is clearly visible.
This leads to different kinematic constraints on the
spectral weight. In the Pfaffian co-propagating
case, for any given $q$ we can make excitations
only within a finite $\omega$ range between
$u_n q$ and $u_c q$.
In contrast,
in the AP case,
the presence of two mutually counter-propagating modes
relevant to each
$A_j$ {\em excludes} the spectral weight from
a finite range of frequencies, at each $q$.\\
\indent {\em MRT conductance.}
We calculate the tunneling current in linear response
using the theory discussed in Ref. \onlinecite{MY1}:
\begin{equation}\label{IVB}
\begin{split}
  I_j(V,\Delta q)\propto \int\,d\omega_1d\omega_2dq_1dq_2\, A_L(\omega_1,q_1) A_j(\omega_2,q_2)\times\\
[f(\omega_1)-
f(\omega_2)] \delta(eV+\omega_1-\omega_2)\delta(\Delta q+ q_1-q_2)\,.
\end{split}
\end{equation}
Here $A_L(\omega,q)$ is the lead spectral function.
We take $A_L(\omega,q)=\delta(\omega-u_Lq)$ corresponding to a $\nu=1$ edge, with $u_L>0$ for the co-propagating lead geometry  and $u_L<0$ for the counter-propagating lead geometry (cf. \Fig{setup}), though other types of leads may be considered.
$f(\omega)$ is the Fermi-distribution function, where we assume zero temperature in the following. $V$ is the applied voltage, and
$\Delta q=ed(B_{||}-B_j)/\hbar c$ is the change in the electron wavevector
relative to the Fermi wavevector, 
where $B_{||}$ is the in-plane magnetic field
and $B_j$ is an offset accounting for different Fermi wavevectors in the lead
and the Pf or AP edge, and $d$ is the distance between the two layers.
$B_j$ is expected to depend on
$j$ as we will further discuss below. This may lead to additional
distinctive features between the Pf and the AP case, since the total current
is the superposition $I(V,B_{||})=\sum_j a_j I_j(V,\Delta q)$ in the latter.
From \Eq{IVB}, it is straightforward to evaluate $I_j(V,\Delta q)$ for
various cases. We present a general result that is valid for any
signs of $u_n$ and $u_L$, and only assumes that $|u_n|$ is smaller
than the ``charged'' velocity parameters $u_c$ and $|u_L|$. We consider
both $u_c<u_L$ and $u_c>u_L$, which leads to qualitative differences in
the co-propagating lead case.
The general result can be glued together from three functions, defined as:
\begin{equation}\label{IABC}
  \begin{split}
  & I_A=\frac{\sgn(u_L)(eV-\Delta qu_L)^2}{(u_L-u_n)(u_c-u_L)^2},
   I_B=\frac{\sgn(u_L)(eV-\Delta qu_n)^2}{(u_L-u_n)(u_c-u_n)^2}\\
  &I_C=\frac{\sgn(u_L)(eV-\Delta qu_c)}{(u_c-u_n)^2(u_c-u_L)^2}\;
(eV(u_n+u_L-2u_c)+\\
&\qquad\qquad\qquad\qquad\qquad \qquad\Delta q(u_nu_c+u_Lu_c-2u_nu_L).
  \end{split}
\end{equation}
For each of these three expressions, we define an associated interval
in $\Delta q$.
Let $J_{AC}$ be the interval between $eV/u_L$ and $eV/u_c$,
$J_B$ the interval between $eV/u_n$ and {\em either}
 $eV/u_L$ {\em or} $eV/u_c$, whichever is closer to $eV/u_n$.
Obviously, $J_{AC}$ and $J_B$ share a common boundary point
and are otherwise disjoint.
\Eq{IABC} was written down with the tacit understanding
that the expressions for $I_A$ and $I_C$ are only valid when
$\Delta q\in J_{AC}$, and that for $I_B$ is only valid for $\Delta q\in J_{B}$.
Outside these intervals, the associated currents are defined to be zero.
For $V>0$ and with these conventions, we find
$I_j=I_A+I_B$ for $u_L(u_L-u_c)u_n>0$ and $I_j=I_C+I_B$ otherwise.
The case $V<0$ is obtained via $I_j(V,\Delta q)=-I_j(-V,-\Delta q)$.\\
\indent{\em Results and discussion.}
\Fig{result} shows our results for $dI_j/dV$ for six cases of interest,
corresponding to the Pf and AP edge state, for co- and counter-propagating
lead geometry, and for both signs of $u_L-u_c$ in the former case.
The most striking difference between the Pf and the AP case is apparent in the co-propagating lead geometry. Here, a positive $V$ requires a positive $\Delta q$ for a current to flow in the Pf case. In contrast, a current will always flow for a range of positive and negative values of $\Delta q$ in the AP case.
These observations are direct consequences of the kinematic constraints on
the spectral function discussed above.
Furthermore, it is only in the Pf co-propagating cases that $dI/dV$ becomes negative.
However,
even for a counter-propagating lead, the Pf and the AP case are clearly
distinguishable.
The smallest mode velocity which is visible in the graph can always
be identified with $u_n$, and its sign distinguishes the Pf from the AP
case.
Also note that in Fig. \ref{result}f) (AP, counter-propagating), $dI/dV$ has no discontinuity {\em within} the region of non-zero current, but does so at one of its boundaries. In contrast, the case in Fig. \ref{result}e) (Pf, counter-propagating) shows a $dI/dV$ discontinuity within the region of non-zero current, but not at its boundaries. Furthermore, even in the AP counter-propagating case (Fig. \ref{result}f) a discontinuity in $d^2I/dV^2$ will clearly distinguish between two different regions (corresponding to $I_j=I_B$ and $I_j=I_C$). The separating line between these two regions has a slope, $u_c$, that differs in sign from the slope $u_L$ of a similar separating line in the Pf counter-propagating case (Fig. \ref{result}e)).
More generally,
Figs. \ref{result}b),c),f) have no discontinuity in $dI/dV$ within the region of non-zero current, but have a discontinuity at one its boundaries, in contrast to the cases in Figs. \ref{result}a),d),e).
Note that in Fig. \ref{result}b) (Pf, co-propagating, $u_c>u_L$), $dI/dV$ smoothly goes through zero {\em within} the region where $I_j=I_C$.
In any case, discontinuities in either $dI/dV$ or $d^2I/dV^2$ allow for a direct measurement of the edge mode velocities.
These findings imply that under all circumstances considered here, the MRT conductance clearly distinguishes the PF edge from the AP edge. This becomes even more pronounced when one takes into account that in the AP case, the MRT current is a superposition of the form 
 $I(V,B_{||})=\sum_j a_j I_j(V,\Delta q)$. For, as discussed above, the contributions $I_j$ do not all feature the same neutral mode velocity $u_n$ in the clean case considered here. Even more importantly, the offset $B_j$ 
entering the definition of $\Delta q$
is expected to depend on $j$ as well. 
This is so because the three operators $\psi_{\text{el},j}$ will in general carry different momenta.
To see this, we may reinterpret these operators in terms of processes taking place at the original $\nu=1$ edge and particle-hole conjugated Pfaffian $\nu=1/2$ edge present in \Eq{LAP}. It is easy to see that, e.g., $\psi_{\text{el},1}$ creates one electron at the $\nu=1/2$ edge while destroying two electrons at the $\nu=1$ edge. Similarly, $\psi_{\text{el},2}$  simply destroys one electron at the $\nu=1$ edge. Hence, if different Fermi momenta are associated with the $\nu=1$ and $\nu=1/2$ components of the edge, all three operators  $\psi_{\text{el},j}$ carry different momenta. In the AP case, one thus expects to measure an MRT conductance which is the superposition of three graphs taken from the appropriate row in the second column of \Fig{result}, with three different horizontal offsets and with two different neutral mode velocities $u_n$.\\
\indent 
We remark that the above results could in principle be affected by edge reconstruction. However, we expect these effects to be considerably weaker
at the second LL edge of interest here. Since this edge is well contained inside
the physical edge of the sample,
fringe field effects, which are usually associated with
edge reconstruction\cite{wan2}, will be weak.
Hence we argue that a picture based
on an unreconstructed $\nu=1/2$ edge may apply.
If edge reconstruction indeed occurs, additional edge modes will result and they can in principle also be detected using the setup discussed here; see Ref. \onlinecite{MY1} for a discussion of this point in the (simpler) context of a $\nu=1/3$ edge, and a detailed study will be left for future work (see also Ref. \onlinecite{overbosch}). We note that we have considered the setup of Fig. \ref{setup} both for its simplicity and experimental relevance\cite{yacoby}; in principle other setups like those of Refs. \onlinecite{kang, yang_i,huber1, huber2} can also be used to study the 5/2 edge.
Finally, we mention a recent alternative proposal to distinguish
the Pf edge from the non-equilibrated AP edge, involving simple two-terminal
measurements \cite{WF}. We caution, however, that the presence of the contacts in such experiments will almost certainly lead to disorder and equilibration among the edge channels, at least near the contacts. As a result the predicted  two-terminal conductance for a scenario based on non-equilibrated edges may never be observed. On the other hand, the (momentum conserving) tunneling 
processes we consider occur away from these contacts, and thus do not suffer from the contact-induced disorder. We are thus hopeful that MRT will prove a useful tool to shed further
light on the
$\nu=5/2$ quantum Hall state in the future.


\begin{acknowledgments}
We are indebted to M. Grayson, W. Kang, C. Nayak, and A. Yacoby
for insightful discussions.
This work was supported by
NSF
grant No. DMR-0907793
 (AS), and NSF
grant No. DMR-0704133 (KY).
\end{acknowledgments}

\end{document}